\documentstyle[preprint,aps]{revtex}

\begin{document}
\preprint{ECT$^*$ 14-98}
\draft
\title
{\bf  Ground State Wave Functions in the Hyperspherical Formalism for
    Nuclei with $A > 4$}
\author{Nir Barnea$^{1)}$, Winfried Leidemann$^{2,3)}$, and
Giuseppina Orlandini$^{2,3)}$}
\address{
1) European Centre for Theoretical Nuclear Physics and Related Areas,
Villa Tambosi, I--38050 Villazzano (Trento), Italy\\
2) Dipartimento di Fisica, Universit\`a di Trento, 
 I--38050 Povo (Trento), Italy\\
3) Istituto Nazionale di Fisica Nucleare, Gruppo collegato di Trento, Italy
}

\date{\today}
\maketitle

\begin{abstract}
The general formulation of a technically
advantageous method to find the ground state solution of the Schr\"odinger
equation in configuration space  for systems with a number of particles
$A$ greater than 4 is presented. 
The wave function is expanded in pair correlated hyperspherical harmonics
beyond the lowest order approximation and then calculated in the Faddeev 
approach. A recent efficient recursive method to construct antisymmetric 
A--particle hyperspherical harmonics is used.
The accuracy is tested for the bound state energies of nuclei
with $A = 6 \div 12$. 
The high quality of the obtained results becomes evident from a comparison 
with other approaches.

\vskip 1truecm
\noindent Keywords: few-body, $A>4$, hyperspherical harmonics.

\end{abstract}
\bigskip

\pacs{PACS numbers: 21.45.+v, 27.20.+n, 31.15.Ja }

\vfill\eject

\section{INTRODUCTION}

Few--body nuclei with a number of nucleons A between 5 and 16  
are a particularly interesting testground for nuclear theory.
They lie in the range between the classical few--nucleon systems ($A \le 4$)
and the smallest nuclei, which can be described realistically
starting from a mean field ansatz. Therefore one hopes that these nuclei could
build a link between few--body and many--body physics. Presently quite an 
effort is made for a better understanding of these intermediate systems. 
Specific interest is devoted to halo--nuclei, but also  the less exotic 
nuclei of this mass range are investigated thoroughly. Many theoretical 
techniques for the calculation of their ground states have been 
imported from the classical few--body field, where there has been a 
considerable
progress in the last years. In fact for the classical few--body systems 
rather different approaches have been developed and proven to lead to precise 
results. These methods include solutions of the Faddeev--Yakubovsky
equation, variational (VMC) and Green Function
Monte Carlo (GFMC), the Hyperspherical Harmonic (HH) ansatz,
the stochastic variational  (SVM), as well as coupled cluster and 
resonating group methods. 

For nuclei with $A>4$ a similar level of precision has not yet been
reached. Here GFMC has led to the most accurate results. 
Exact bound state energies with realistic $NN$ interactions have been 
calculated for $A \le 9$ ~\cite{PPCPW98}. 
Unfortunately wave functions cannot be generated with this method. Recently, 
a very powerful tool to calculate few--body wave functions has been developed 
with the SVM ansatz ~\cite{{Varga95},{Varga96}}.
However, this approach is probably most suitable for 
systems with $A < 8$. For nuclei with $A \ge 8$ rather good results have been 
obtained in the Integro--Differential Equation Approach (IDEA)
~\cite{Brizzi96}, which uses the 
HH expansion at first order, and with a variational method, the Translational
Invariant Configuration Interaction Method (TICI) ~\cite{{BFBBG90},{GMNBPW96}},
which is inspired by the coupled cluster method.

In this work we present the general formulation of the method, which combining
the main ideas of the HH expansion, the pair correlation ansatz and the 
Faddeev approach allows to calculate wave functions of few--body systems.
We apply it to calculate the binding energies of
nuclear systems with 6, 8 and 12 particles.
Including higher order HH functions we make the 
nontrivial step beyond the IDEA approach.
The difficulty in constructing antisymmetric $A$--particle HH is overcome
by the use of a recently developed very efficient recursive method 
~\cite{Nir97}, where HH basis functions are constructed, 
belonging to well defined
irreducible representations of the orthogonal and symmetric groups.

The paper is organized as follows. The method is described in Section 
\ref{gen}, while the construction of the HH basis function is
briefly reviewed in Section \ref{basis}. Section \ref{matrix} illustrates
how the matrix elements are calculated and the obtained results for the 
binding energies of nuclei with $A = 6, 8$ and 12 are discussed in
Section \ref{results}. Conclusions are drawn in Section \ref{concl}.

\section{GENERAL FORMULATION OF THE METHOD}\label{gen}

Our aim is the solution of the Schr\"odinger equation for a system of 
A particles interacting via a two--body potential. After subtraction of the
center of mass Hamiltonian the problem reduces to the solution of the 
internal Hamiltonian in terms of $N=(A-1)$ Jacobi coordinates $\vec\eta_i$.
One can  write the wave function in terms of an  HH expansion 
\begin{eqnarray} \label{HH}
    \Phi(\vec\eta_1..\vec\eta_{A-1}, s_1..s_A, t_1..t_A)&=&\sum_{K\,\nu} 
              R_{K\,\nu}(\rho) H_{K\,\nu}(\Omega,s_1..s_A, t_1..t_A)\cr
    &\equiv& \sum_{K\nu}\Phi_{K\nu}(\vec\eta_i,s_1..s_A, t_1..t_A)\,,
\end{eqnarray}
where, in case of nucleons, 
\begin{eqnarray} \label{H}
  H_{K\,\nu}(\Omega,s_1..s_A, t_1..t_A) = 
         \sum_{Y_{A-1}}
         \frac{\Lambda_{\Gamma_{A},Y_{A-1}}}{\sqrt{| \Gamma_{A}|}} \,
         &{\cal Y}&_{K_N L_N M_N
          \, \Gamma_{A} Y_{A-1}\, \alpha^K_N}(\Omega)\,\times\cr
          &{\cal X}&_{S S^z T T^z \, \tilde{\Gamma}_{A},\tilde Y_{A-1}
         \, \alpha^{ST}_A}(s_1..s_A, t_1..t_A)
\end{eqnarray}
represent the hyperspherical harmonic functions coupled to the spin--isospin 
basis functions to yield the totally antisymmetric wave function.
The ${\cal Y}_{K_N L_N M_N \, \Gamma_{A}Y_{A-1} \, \alpha^K_N}(\Omega)$ 
are the so called symmetrized hyperspherical functions that depend
on the $N$ Jacobi coordinates.
They are
HH functions with hyperspherical angular momentum $K=K_N$, and good angular
momentum quantum numbers $L_N, M_N$ that belong
to well defined irreducible representations (irreps) 
$\Gamma_{1} \in \Gamma_2 \ldots \in \Gamma_A $, of the permutation 
group--subgroup chain  
${\cal S}_1 \subset {\cal S}_2 \ldots \subset {\cal S}_A $,
denoted by the Yamanouchi symbol 
$[ \Gamma_A, Y_{A-1} ] \equiv [ \Gamma_A,\Gamma_{A-1},\ldots,\Gamma_1 ]$.
The dimension of the irrep $\Gamma_{n}$ is denoted by $| \Gamma_{n}|$
and $\Lambda_{\Gamma_{A},Y_{A-1}}$ is a phase factor \cite{Akiva88}.
Similarly, the function 
${\cal X}_{S S^z T T^z \, \tilde{\Gamma}_{A}\tilde{Y}_{A-1}
         \, \alpha^{ST}_A}$
is a symmetrized spin--isospin state with good quantum numbers $S, S^z, T$
and $T^z$.
The label $\alpha^K_N$ ($\alpha^{ST}_A$) is needed to remove the degeneracy 
of the hyperspherical (spin--isospin) states with a given symmetry.
The argument of the hyperradial function  $R_{K\,\nu}$
is the hyperradius $\rho$ defined by
\begin{equation}
\rho^2=\sum_i \eta_i^2\,,
\end{equation}
$\Omega$ is the (3A-4)--dimensional hyperangle, and for brevity we shall
use $\nu$ for all the quantum numbers but $K$, i.e. 
$\nu \equiv (L_N M_N \, S S^z T T^z \, \Gamma_A \alpha^K_N \alpha^{ST}_A)$.
In what follows we shall use the subscript $N$ to denote $A$ particle
quantum numbers that depend on the Jacobi coordinates, and the subscript 
$A-1$ for the $A-1$ particle subsystem.

It is well known that finding a solution of the Schr\"odinger equation in
terms of the ansatz (\ref{HH})
can be very difficult because the number of basis functions increases very 
fast with $K$ and in order to 
have a real convergence one must use a huge number of basis functions
\cite{Kievsky97}. 
Therefore a correlation function is advantageous to give the wave function 
a proper behavior ~\cite{FE72}. 
Its advantages have been extensively verified for classical few-body
systems reaching a high level of accuracy \cite{Kievsky93,Victor89}.
A general ansatz within the two--body correlation scheme is the Jastrow 
factor
\begin{equation} \label{Jastrow}
   \Psi = \prod_{i<j} f_{ij} \Phi\,,
\end{equation}
where $f_{ij}$ is a two--body correlation function.
However, the use of the Jastrow ansatz leads to $3A-3$
dimensional integrals. Therefore it is more convenient to use the so called
pair correlation ansatz 
\begin{equation} \label{Psi}
\Psi=\sum_{i<j} \chi_{ij} \Phi,
\end{equation}
because in this case one can use the Faddeev approach which leads to at
most four--body integrals.
In the Faddeev approach the Schr\"odinger equation is replaced by 
equivalent equations, 
\begin{equation} \label{Faddeev}
   (T-E)\Psi_{ij}=-V_{ij} \Psi \; ,
\end{equation}
where $ \Psi = \sum_{i<j} \Psi_{ij}$. 
In order to speed up the convergence these equations can be further 
modified \cite{Brizzi96} to include
the contribution of the hypercentral potential explicitly.
The hypercentral potential $V_{hc}(\rho)$ is defined as the projection 
of the two--body
interaction on the subspace of the lowest order hyperspherical state, i.e.
the hyperspherical state expressed in  Eq. (\ref{H}), with the minimal
$K$:
\begin{equation} \label{Vhc}
  V_{hc}(\rho)=
        \int d \Omega \, H_{K_{min} \nu}^{\dagger}(\Omega,s_i,t_i) 
         V(\vec{r}_{1,2},s_1,s_2,t_1,t_2) 
         H_{K_{min}\nu} (\Omega,s_j,t_j)\,. 
\end{equation}
Here the integration is carried over the $3N-1$ dimensional hypersphere,
and  an implicit summation over all the spin--isospin states is understood.
With the help of this definition we can rewrite Eq. (\ref{Faddeev}) as
follows
\begin{equation} \label{Fad_hc}
   \left[ T+\frac{A(A-1)}{2}V_{hc}(\rho)-E \right]\Psi_{ij}=
   -[V_{ij}-V_{hc}(\rho)] \Psi \; .
\end{equation}
Motivated by the pair--correlation ansatz (\ref{Psi}) we shall expand
the Faddeev amplitude $\Psi_{ij}$ in the following way,
\begin{equation} \label{Psi_ij}
  \Psi_{ij}=\sum_{p} \Phi_p (\rho,\Omega) \chi_p(z_{ij}) 
	    =\sum_{K \nu p} R_{K\nu p}(\rho) 
               H_{K\nu} (\Omega,s_1..s_A, t_1..t_A)\chi_p(z_{ij}) \,,
\end{equation}
where $\chi_p(z_{ij})$ is a polynomial of order $p$, and 
$z_{ij}$ is related to the relative two--body distance through
\begin{equation}
	z_{ij} = r_{ij}/\rho \,.
\end{equation}
Although expanding the correlation function in terms of $z_{ij}$
should be equivalent to an expansion in $r_{ij}$, it is more stable 
numerically to take $\chi_p(z_{ij})$ since $r_{ij}$ goes to zero with $\rho$.
Substituting the expansion of Eq. (\ref{Psi_ij}) into Eq. (\ref{Faddeev})
we get 
\begin{equation} \label{Fad2}
   \left[-{1 \over 2}\sum_{n=1}^{A-1} \Delta_n
   +\frac{A(A-1)}{2}V_{hc}(\rho) -E \right]
   \sum_{p} \Phi_{p} \chi_p(z_{ij}) 
   =-[V_{ij}-V_{hc}(\rho)] \sum_{p} \Phi_{p} \sum_{k>l} \chi_p(z_{kl}) \,,
\end{equation}
where $\Delta_n$ is the Laplace operator associated with the $n$'th Jacobi
coordinate $\vec{\eta}_n$.
Multiplying  Eq. (\ref{Fad2}) on the left by 
$( H_{K'\nu'} \chi_{p'}(z_{ij}) )^{\dagger}$ and using the notation
\begin{equation} \label{defME}
   _{ij}< K' \nu' p' | \hat O | K \nu p >_{kl} = 
   \int d \Omega \, H_{K'\nu'}^{\dagger}(\Omega,s_n,t_n)
   \chi_{p'}^*(z_{ij})
   \hat O H_{K \nu}(\Omega,s_m,t_m) \chi_p(z_{kl})
\end{equation}
we get 
\begin{eqnarray} \label{Fad3}
 & & \sum_{K \nu p} \,
    _{ij}< K' \nu' p' |-\frac{1}{2}\sum_{n=1}^{A-1} \Delta_n 
    +\frac{A(A-1)}{2}V_{hc}(\rho) - E  | K \nu p >_{ij}
    R_{K \nu p}(\rho) = \nonumber \\ & & \hskip 2truecm  
   -\sum_{K'' \nu'' p''} \, _{ij}< K' \nu' p' | [V_{ij}-V_{hc}(\rho)]
    \sum_{k>l}| K'' \nu'' p'' >_{kl} R_{K'' \nu'' p''}(\rho) \; .
\end{eqnarray}
If the norm matrix
\begin{equation}
  A_{K' \nu' p', K \nu p} \equiv 
     _{ij}< K' \nu' p' | K \nu p >_{ij}
\end{equation}
is non singular, one can define the projection operator
\begin{equation}
  \hat Q = \sum_{K \nu p, K' \nu' p' } | K' \nu' p' >
           A_{K' \nu' p', K \nu p}^{-1} < K \nu p | \,.
\end{equation}
The projection operator commutes with $V_{ij}$ and
one can rewrite Eq. (\ref{Fad3}) in the following form
\begin{eqnarray} \label{Fad-A}
 & & \sum_{K \nu p} \,
    _{ij}< K' \nu' p' |-\frac{1}{2}\sum_{n=1}^{A-1} \Delta_n 
    +\frac{A(A-1)}{2}V_{hc}(\rho)    - E  | K \nu p >_{ij}
    R_{K \nu p}(\rho) = \nonumber \\ & & \hskip 2truecm - 
    \sum_{K \nu p} \sum_{K'' \nu'' p''} \sum_{K''' \nu''' p'''}
    \, _{ij}< K' \nu' p' | [V_{ij}-V_{hc}(\rho)] | K'' \nu'' p'' >_{ij} 
    \times	\nonumber \\ & & \hskip 2truecm
    A_{K'' \nu'' p'', K''' \nu''' p'''}^{-1}
    \sum_{kl} \, _{ij}< K''' \nu''' p''' | K \nu p >_{kl} 
    R_{K \nu p}(\rho) \; .
\end{eqnarray}
Using the hyperspherical coordinates, and the following definitions
\begin{eqnarray}
  T_{K'\nu'p',K\nu p} &=&
        _{ij}< K' \nu' p' | \hat K | K \nu p >_{ij}\,, \cr
  V^r_{K'\nu'p',K\nu p}(\rho) &=&
        _{ij}< K' \nu' p' | V_{ij}-V_{hc} | K \nu p >_{ij}\,,\cr
  W^{[3]}_{K'\nu'p',K\nu p} &=& \sum_{K'' \nu'' p''}
        A_{K' \nu' p', K'' \nu'' p''}^{-1} \,
        _{ij}< K'' \nu'' p'' | K \nu p >_{ik}\,,
        \hskip 2truecm k \neq i,j \,,\cr
  W^{[4]}_{K'\nu'p',K\nu p} &=& \sum_{K'' \nu'' p''}
        A_{K' \nu' p', K'' \nu'' p''}^{-1} \,
        _{ij}< K'' \nu'' p'' | K \nu p >_{kl}\,,
        \hskip 2truecm k,l \neq i,j \,,
\end{eqnarray}
Eq. (\ref{Fad-A}) can be expressed as an hyperradial equation
\begin{eqnarray} \label{radial}
  \sum_{\mu' } \Big[ &-& \frac{1}{2} A_{\mu \mu'}
  \big( \frac{d^2}{d\rho^2} + \frac{3 A -4}{\rho}\frac{d}{d\rho} 
  \big) +\frac{1}{2}\frac{T_{\mu \mu'}}{\rho^2}
   +\frac{A(A-1)}{2}V_{hc}(\rho)A_{\mu \mu'}
   -E \, A_{\mu \mu'} \Big] R_{\mu'}(\rho) \cr
  &=&-\sum_{\mu' \mu''} V^r_{\mu \mu'} (\rho)
  \left( \delta_{\mu'\, \mu''}
  +2(A-2)W^{[3]}_{\mu', \mu''}
  +\frac{(A-2)(A-3)}{2} W^{[4]}_{\mu',\mu''}
  \right)  R_{\mu''}(\rho) \,.
\end{eqnarray}
Here $\mu \equiv (K \nu p)$, and $\hat K$ is the generalized,
hyperspherical, angular momentum operator.
This hyperradial equation can be solved by expanding the 
hyperradial function $R_{\mu}(\rho)$ into basis functions 
$\phi_{n_{\rho}}(\rho)$
\begin{equation}
   R_{\mu}(\rho) = \sum_{n_{\rho}} 
                   C_{n_{\rho}}^{\mu} \phi_{n_{\rho}}(\rho)
\end{equation}
containing the Laguerre polynomials $L_{n_{\rho}}^a$
\begin{equation} \label{Frho}
   \phi_{n_{\rho}}(\rho) = \sqrt{n_{\rho}! \over (n_{\rho}+a)!} \,\,
    b^{-3({\it A}-1)\over 2}
    \left(\frac{\rho}{b}\right)^{{\it a}-(3 {\it A}-4)\over 2}
    L_{n_{\rho}}^a\left({\rho\over b}\right)\,
    \exp\left[-{\rho\over 2b}\right]\,.
\end{equation}
Multiplying the right and left hand sides by $\phi^\star_{n_{\rho}}(\rho)$ 
and performing the hyperradial integration one remains with
the generalized eigenvalue problem
\begin{equation}
\sum_{m} C_m ({\bf H}_{m n}-E {\bf M}_{m n})=0\,,
\end{equation}
where $m$ and $n$ stand for the sets ${K,\nu,p,n_{\rho}}$ and 
${K',\nu',p',n_{\rho}'}$, respectively,
and the matrix elements ${\bf H}_{m n}$ and ${\bf M}_{m n}$ are
given by the corresponding hyperradial integrals.

Eq. (\ref{radial}) represents the main equation of the present method. 
The following remarks have to be stressed in order here:

\begin{enumerate}

\item the 3(A-1)--dimensional Schr\"odinger equation has been reduced to
      a $3\times 4$ dimensional integro--differential equation, due to the
      presence of at most four--body terms ($W^{[4]}_{K'\nu'K\nu}$);

\item to the extent the convergence in the expansions is reached the result 
      for $\Psi=\sum_{i<j} \Psi_{ij}$ can be considered
      as an "exact" ground state solution of the many--body 
      Schr\"odinger equation.
\end{enumerate}
\section{CONSTRUCTION OF THE BASIS FUNCTIONS}\label{basis}

Technical difficulties in calculating the matrix elements 
${\bf H}_{m n}$ and ${\bf M}_{m n}$ are encountered in the construction
of the  basis functions $H_{K\,\nu} (\Omega,s_i,t_i)$ for increasing numbers of
particles and hyperspherical angular momentum $K$, as well as in the
calculation of the two--, three-- and four--body terms that appear in 
${\bf H}_{m n}$ and ${\bf M}_{m n}$.
The basis functions $H_{K\,\nu} (\Omega,s_i,t_i)$ must be totally
antisymmetric functions in total space (coordinate, spin, isospin). 
Therefore, one needs efficient
algorithms to construct convenient expressions for the symmetrized
spin--isospin basis functions as well as symmetrized hyperspherical
wave functions. 

\subsection{Spin--isospin states with arbitrary permutational symmetry}

The construction of A--particle symmetrized spin--isospin
functions as well as hyperspherical functions
can be done recursively \cite{Akiva88}.
Each  A--particle spin--isospin function of a well defined 
irrep of the symmetric group 
$\Gamma_A$ and spin--isospin quantum numbers $S_A, T_A$,
is written as a linear combination of spin--isospin coupled products 
of an (A-1)--particle wave function of well defined irrep $\Gamma_{A-1}$
and spin--isospin quantum numbers $S_{A-1}, T_{A-1} $,
and the A'th particle wave function with spin--isospin quantum numbers
$s_A, t_A$. 
The coefficients of this
linear combination are a sort of coefficients of fractional parentage.
The subspace of functions with good quantum numbers $S_A, T_A$ and
$\Gamma_{A-1}$ is an invariant subspace for the transposition class--sum
operator of the symmetric group ${\cal S}_A$. The eigenvalues that are obtained
after the diagonalization of the class--sum operator identify
the irreps of the symmetric group uniquely. The eigenvectors are the 
spin--isospin coefficients of fractional parentage (stcfps)
\begin{eqnarray} \label{stcfps}
& & {\cal X}_{S_A {S^z}_A T_A {T^z}_A \, 
    {\Gamma}_{A} Y_{A-1}  \, \alpha^{ST}_A} (s_i t_i) 
 =  \sum_{S_{A-1} T_{A-1} \alpha^{TS}_{A-1} }
\nonumber \\ & & \hskip 2truecm
   [ (S_{A-1}; s_A)S_A (T_{A-1} ; t_A)T_A \Gamma_{A-1}\alpha^{ST}_{A-1}
   |\} S_A T_A {\Gamma}_{A} \alpha^{ST}_A ] \times
\nonumber \\ & & \hskip 2truecm
  \big[ {\cal X}_{S_{A-1} {S^z}_{A-1} T_{A-1} {T^z}_{A-1} \, 
  {\Gamma}_{A-1} Y_{A-2}
  \, \alpha^{ST}_{A-1} } \otimes s_{A} t_A \big]^{S_A {S^z}_A T_A {T^z}_A}\,.
\end{eqnarray}
The sum over $S_{A-1}$ and $T_{A-1}$ in Eq. (\ref{stcfps}) is subject to the 
usual angular momentum coupling rules.

\subsection{Symmetrized hyperspherical functions}

In an analogous way, one can use the recursive methods
developed in the last few years  \cite{{Akiva94},{Nir98b}} for
constructing hyperspherical functions that belong to well defined 
irreps of the symmetric group.
In this methods the reduction problem 
${\cal O}_{3A-3} \supset {\cal O}_3 \otimes {\cal S}_A$ is solved and one obtains
hyperspherical functions
which belong to irreps of the symmetric group ${\cal S}_A$ and have 
good angular momentum and hyperspherical angular momentum quantum numbers.
These functions are expressed in terms of ${\cal S}_{A-1}$ hyperspherical states,
that are coupled, via the ``tree" method  
 ~\cite{Vilenkin66}, 
with the appropriate single--Jacobi coordinate hyperspherical functions, 
into coupled $A$--particle states with the desired
angular momentum and hyperspherical angular momentum quantum numbers.
Actually, these coupled states yield an invariant subspace with 
respect to ${\cal S}_A$. The transposition class--sum of the symmetric 
group (the second Casimir operator) is diagonalized within this subspace.
The eigenvalues that are obtained after the diagonalization  
identify the irreps of the symmetric group uniquely, and
the eigenvectors are the hyperspherical coefficients of fractional 
parentage (hscfps). 

Constructing basis functions in such a recursive way makes the evaluation
of any two body operator easy, since only the matrix element of the two--body
operator between the last two particles needs to be calculated.
A further improvement in the efficiency of the algorithm 
can be reached if one uses reversed order Jacobi coordinates for
the construction of the hyperspherical functions.
Normally Jacobi coordinates are defined so that the first
one is the relative distance between particles 1 and 2, while the last
is the distance between the A'th particle and the center of mass of the
A-1 particle system. This implies that in order to calculate the matrix
element of a two--body operator, depending typically 
on the interparticle distance between the A'th and (A-1)'th particles (see
Section \ref{matrix}),
one needs to rotate the last two Jacobi coordinates by a proper angle so 
that one of them represents the interparticle distance. This can be done
by using the Raynal Revai, the T-coefficients and the 6-j coefficients. 
Constructing the Jacobi coordinates in reverse order, would
simplify the calculation of two-- and three--body matrix elements as no further
rotation is needed \cite{Nir98b}.

Recently an alternative way to construct the symmetrized
hyperspherical functions has been 
proposed \cite{Nir97}. This method consists in introducing as an 
intermediate subgroup the orthogonal group of kinematic rotations 
${\cal O}_{A-1}$,
i.e. one uses the group chain 
${\cal O}_{3A-3}\supset {\cal O}_3 \otimes {\cal O}_{A-1}\supset 
{\cal O}_3 \otimes {\cal S}_A$.
The introduction of the kinematic group turns out to be of particular
importance for increasing number of particles and for large values of $K$
as it results in a significant reduction in the number of hscfps and in
the computation time. 
Another benefit in using this method is the realization of kinematic
rotations through the representation matrices of the group ${\cal O}_{A-1}$
thus avoiding the use of the Raynal--Revai and the T--coefficients
in calculating matrix elements of two--body operators depending on the 
interparticle distance.
In the present calculation we have adopted this procedure, even if it is
less efficient than defining Jacobi coordinates in reversed order. The 
reason is that in this way one can easily extend the present method
to solve the Schr\"odinger--like equation with source, necessary for
the calculation of the Lorentz integral transforms of response functions ~\cite{ELO94}. 
The presence of the source annuls the advantages of the alternative set of
Jacobi coordinates.

Summarizing the procedure for the construction of the symmetrized 
hyperspherical harmonics one can split it into two steps. 
In the first step one constructs
hyperspherical functions with good orthogonal symmetry, i.e. basis
functions with good quantum numbers $K_N, L_N, M_N$ that belong
to a well defined Gel'fand--Zetlin pattern 
$\mbox{\boldmath $\Lambda$}_{N}= 
	 (\mbox{\boldmath $\lambda$}_{N},
          \mbox{\boldmath $\lambda$}_{N-1},\ldots,
          \mbox{\boldmath $\lambda$}_{2})$.
The orthogonal group irreps $\mbox {\boldmath $\lambda$}_j$,($j=2,3,\ldots $)
are characterized by the integer or half integer
numbers
$\lambda_{j,1}$, $\lambda_{j,2}$, $ \ldots $ , $\lambda_{j,k}$, 
where $k=[\frac{j}{2}]$. 
In Ref. \cite{Nir97} it was pointed out that for hyperspherical 
functions these 
numbers are always integers and, in addition, there are at 
most three non--zero 
values in the irrep \mbox{\boldmath $\lambda$}$_j$ for $j \geq 6 $ i.e., 
$\lambda_{j,1}$, $\lambda_{j,2}$, and $\lambda_{j,3}$.
The second step is to reduce each irrep of the orthogonal group 
${\cal O}_N$ into
irreps of the symmetry group ${\cal S}_{N+1}$. 
These two steps are carried out using the recursive method developed
in Ref. \cite{Nir97}. 
At the end, the symmetrized A--particle, 
$N$--Jacobi coordinate, hyperspherical states can be expressed in terms
of $(A-1)$--particle states coupled to the $A$'th particle state by using
two new types of coefficients of 
fractional parentage, namely the orthogonal--hyperspherical cfps (ohscfps)
for the construction of hyperspherical functions with good orthogonal 
symmetry and the orthogonal cfps (ocfps) for the reduction 
${\cal O}_{A-1} \downarrow {\cal S}_{A}$.
One has
\begin{eqnarray} \label{symhh}
\label{twocfp} & &
{\cal Y}_{ 
  K_{N} L_{N} M_{N} \mbox {\boldmath $\lambda$}_{N} \beta^K_{N} 
  \Gamma_A Y_{A-1}  \beta^{\lambda}_A} (\Omega)
\nonumber \\ & &
= \sum_{ {\tiny \mbox {\boldmath $\lambda$}_{N-1}} \beta^{\lambda}_{A-1} }
 [ \left( \mbox {\boldmath $\lambda$}_{N-1} \Gamma_{A-1} 
 \beta^{\lambda}_{A-1} \right) 
   \mbox{\boldmath $\lambda$}_{N} 
  |\} \mbox {\boldmath $\lambda$}_{N} \Gamma_A \beta^{\lambda}_A ]\times 
\nonumber \\ & &
\sum_{K_{N-1} L_{N-1} \beta^K_{N-1}  \ell_{N} } 
  \left[  \big( K_{N-1} L_{N-1} \mbox {\boldmath $\lambda$}_{N-1} 
  \beta^K_{N-1};\ell_{N} \big)
  K_{N} L_{N} |\} K_{N} L_{N} \mbox {\boldmath $\lambda$}_{N} 
\beta^K_{N} \right]
 \nonumber \\ & & \; \; \;
{\cal Y}_{
   \big( K_{N-1} L_{N-1}  \mbox {\boldmath $\lambda$}_{N-1} \beta^K_{N-1}
   \Gamma_{A-1} Y_{A-2}   \beta^{\lambda}_{A-1} ; \ell_{N} \big) 
   K_{N} L_{N} M_{N} }(\Omega) \; \; .
\end{eqnarray}
Note that the order of the summation
in this equation is important. We have to start from the sum over the 
ocfps
\begin{equation}
[ \left( \mbox {\boldmath $\lambda$}_{N-1} \Gamma_{A-1} 
 \beta^{\lambda}_{A-1} \right)  \mbox{\boldmath $\lambda$}_{N} 
  |\} \mbox {\boldmath $\lambda$}_{N} \Gamma_A \beta^{\lambda}_A ] \, ,
\nonumber
\end{equation} 
to determine the irrep $\mbox{\boldmath $\lambda$}_{N-1}$ of ${\cal O}_{N-1}$ 
and then we can sum over the ohscfps
\begin{equation} \label{cfpON}
\left[  \big( K_{N-1} L_{N-1} \mbox {\boldmath $\lambda$}_{N-1} 
  \beta^K_{N-1};\ell_{N} \big)  K_{N} L_{N} |\} K_{N} L_{N}
  \mbox {\boldmath $\lambda$}_{N} \beta^K_{N} \right] \, ,
\nonumber
\end{equation}
where $(\beta^K_N$ and $\beta^{\lambda}_A)$ are the degeneracy removing 
labels ($(\beta^K_N,\beta^{\lambda}_A)\equiv\alpha^K_N$). 
 
\section{Calculation of the matrix elements}\label{matrix}

The matrix elements of any two--body operator
\begin{equation} \label{O_ij}
   O_{ij} = O^R(\vec{r}_{ij}) O^{ST}(s_i,s_j,t_i,t_j)
\end{equation}
between the fermionic, antisymmetric, hyperspherical basis functions of
Eq. (\ref{H}), can be written as a sum of a spatial term multiplied by
a spin--isospin term,
\begin{eqnarray} \label{2bme}
 & & < K \nu | O_{ij} | K' \nu' > =  
      \sum_{\Gamma_{A-1}\Gamma'_{A-1}\Gamma_{A-2}}
      \prod (\mbox{ ISF }) 
      \times \cr & & \hskip 2truecm
     < S_A {S^z}_A T_A {T^z}_A \, \tilde{\Gamma}_{A} \tilde{Y}_{A-1}
        \, \alpha^{ST}_A |   O^{ST}(s_i,s_j,t_i,t_j)    
      | S'_A {S'^z}_A T'_A {T'^z}_A \, \tilde{\Gamma}'_{A} \tilde{Y}'_{A-1} 
        \, \alpha'^{ST}_A >
      \times \cr & & \hskip 0.5truecm
      < K_{N} L_{N} M_{N} \mbox {\boldmath $\lambda$}_{N} 
       \beta^K_{N} \Gamma_A Y_{A-1}  \beta^{\lambda}_A |
        O^R(\vec{r}_{ij})
      | K'_{N} L'_{N} M'_{N} \mbox {\boldmath $\lambda$}'_{N} 
       \beta'^K_{N} \Gamma'_A Y'_{A-1}  \beta'^{\lambda}_A > \,,
\end{eqnarray}
where the sum runs over all the ${\cal S}_{A-2}, {\cal S}_{A-1}$ irreps 
$\Gamma_{A-1}\Gamma'_{A-1}\Gamma_{A-2}$ such that
$\Gamma_{A-2}\in\Gamma_{A-1}\in\Gamma_{A}$ and 
$\Gamma_{A-2}\in\Gamma'_{A-1}\in\Gamma'_{A}$.
The factor preceding the spatial and spin--isospin matrix element is
a product of the inner product symmetric group isoscalar factors for
the antisymmetric representation and is given by
\begin{equation}
 \prod( \mbox{ ISF }) = 
    \Lambda_{\Gamma_{A}\Gamma_{A-1}}
         \sqrt{\frac{|\Gamma_{A-1}|}{|\Gamma_{A}|}}
    \Lambda_{\Gamma_{A-1}\Gamma_{A-2}}
         \sqrt{\frac{|\Gamma_{A-2}|}{|\Gamma_{A-1}|}}
    \Lambda_{\Gamma'_{A}\Gamma'_{A-1}}
         \sqrt{\frac{|\Gamma'_{A-1}|}{|\Gamma'_{A}|}}
    \Lambda_{\Gamma'_{A-1}\Gamma'_{A-2}}
         \sqrt{\frac{|\Gamma'_{A-2}|}{|\Gamma'_{A-1}|}}\, .
\end{equation}
The phase factor
 $\Lambda_{\Gamma_{A}  \Gamma_{A-1}}$ \cite{Akiva88b}
is positive (negative) when the number of boxes in $\Gamma_{A}$
below the row of the $A$'th particle is even (odd).

Using the permutation symmetry of the HH states the matrix elements of 
any two--body operator, $O_{ij}$, are equal to the matrix elements of 
$O_{A \,,A-1}$. The spatial matrix elements are then calculated
using the last two generations of cfps, referring to the 
constructions $ A-2 \longrightarrow A-1$ and
$ A-1 \longrightarrow A $,
and the appropriate kinematical rotations which
reduce the calculation of any
two--body operator to sum over one dimensional integrals.
The formal derivation of the calculation 
of two--body operator matrix elements between
the symmetrized hyperspherical harmonics of Eq. (\ref{symhh}) has been
presented in rather a detailed manner by Barnea and Novoselsky 
\cite{Nir98a}. Summarizing their results one has 
\begin{eqnarray} \label{DVN}
 & & < K_{N} L_{N} M_{N} \mbox {\boldmath $\lambda$}_{N} 
       \beta^K_{N} \Gamma_A Y_{A-1}  \beta^{\lambda}_A |
        O^R(\vec{r}_{A,A-1})
      | K'_{N} L'_{N} M'_{N} \mbox {\boldmath $\lambda$}'_{N} 
       \beta'^K_{N} \Gamma'_A Y'_{A-1}  \beta'^{\lambda}_A > =
\cr & & \hskip 2truecm
	\delta_{\Gamma_{A-2}, \Gamma'_{A-2}}
	\sum_{\mbox {\boldmath $\lambda$}_{N-1} \beta^{\lambda}_{A-1}}
	\sum_{\mbox {\boldmath $\lambda$}'_{N-1} 
              \beta^{\lambda \, \prime}_{A-1}}
	\sum_{\mbox {\boldmath $\lambda$}_{N-2} \beta^{\lambda}_{A-2}}
\cr & & \hskip 2truecm
    [ \left( \mbox {\boldmath $\lambda$}_{N-1} \Gamma_{A-1} 
      \beta^{\lambda}_{A-1} \right)  \mbox{\boldmath $\lambda$}_{N} 
      |\} \mbox {\boldmath $\lambda$}_{N} \Gamma_A \beta^{\lambda}_A ]^* 
    [ \left( \mbox {\boldmath $\lambda$}'_{N-1} \Gamma'_{A-1} 
      \beta^{\lambda \, \prime}_{A-1}\right) \mbox{\boldmath $\lambda$}'_{N} 
    |\} \mbox {\boldmath $\lambda$}'_{N} \Gamma'_A \beta^{\lambda\,\prime}_A ]
\cr & & \hskip 1truecm \times
    [ \left( \mbox {\boldmath $\lambda$}_{N-2} \Gamma_{A-2} 
      \beta^{\lambda}_{A-2} \right)  \mbox{\boldmath $\lambda$}_{N-1} 
    |\} \mbox{\boldmath $\lambda$}_{N-1}\Gamma_{A-1}\beta^{\lambda}_{A-1}]^* 
    [ \left( \mbox {\boldmath $\lambda$}_{N-2} \Gamma_{A-2} 
      \beta^{\lambda}_{A-2} \right)  \mbox{\boldmath $\lambda$}'_{N-1} 
    |\} \mbox{\boldmath $\lambda$}'_{N-1}\Gamma'_{A-1}
        \beta^{\lambda\,\prime}_{A-1}]
\cr & & \hskip 2truecm \times 
    \sum_{\mbox {\boldmath $\lambda$}''_{N-1}}
    \mbox { $D$ }_{ \tilde{\bf \Lambda}_{N-1}^{ \prime \prime }
                    \tilde{\bf \Lambda}_{N-1} }
                 ^{  \mbox {\boldmath $\lambda$}_{N} \; * }
                  (g_{A,A-1})
    \mbox { $D$ }_{ \tilde{\bf \Lambda}_{N-1}^{ \prime \prime } 
                    \tilde{\bf \Lambda}'_{N-1} }
                 ^{  \mbox {\boldmath $\lambda$}'_{N} }
                  (g_{A,A-1}) 
\cr & & \hskip 1truecm \times
    < K_{N} L_{N} M_{N} \mbox {\boldmath $\lambda$}_{N} \beta^K_{N}   
    \tilde{\bf \Lambda}^{\prime \prime} _{N-1} 
    Y_{A-2}  \beta^{\lambda}_{A-2}) 
    |  O^R (\sqrt{2} \vec{\xi}_{N} ) 
    | K'_{N} L'_{N} M'_{N} \mbox {\boldmath $\lambda$}'_{N} 
    \beta^{K \, \prime}_{N} \tilde{\bf \Lambda}^{ \prime \prime}_{N-1} 
    Y_{A-2}  \beta^{\lambda}_{A-2}) > \; .
\end{eqnarray}
Here, for simplicity, we use the symbol ${\bf \Lambda}_{N-1} $ 
to denote the irreps 
$ [\mbox {\boldmath $\lambda$}_{N-1}, \mbox {\boldmath $\lambda$}_{N-2}]$.
Note that the sum in Eq. (\ref{DVN}) over the irreps 
$ \mbox {\boldmath $\lambda$}^{\prime \prime}_{N-1} $ is restricted to those 
states which belong both to the irrep $ \mbox {\boldmath $\lambda$}_{N}$
and $ \mbox {\boldmath $\lambda$}'_{N}$.
The kinematical ``relative coordinate'' rotation applied to the Jacobi coordinates 
\begin{eqnarray} \label{rotation1}
\vec{\xi}_{N-1}  & = & \sqrt{\frac{2 (A-2)}{A}} 
                       \big( \frac{\vec{r}_{A-1}+\vec{r}_{A}}{2}
			  -\frac{1}{A-2}\sum_{i=1}^{A-2} \vec{r}_i \big)
\nonumber \\ 
\vec{\xi}_{N}    & = & \sqrt{\frac{1}{2}}
                       \big( \vec{r}_{A}-\vec{r}_{A-1} \big)
                 \; ,                     
\end{eqnarray} 
and given by
\begin{eqnarray} \label{rotation2}
\vec{\xi}_{N-1}  & = & \sqrt{\frac{A}{2(A-1) } }\vec{\eta}_{N-1}  +
                       \sqrt{\frac{A-2}{2(A-1)}}\vec{\eta}_{N}
\nonumber \\ 
\vec{\xi}_{N}    & = & -\sqrt{\frac{A-2}{2(A-1)}}\vec{\eta}_{N-1}
                       +\sqrt{\frac{A}{2(A-1) } }\vec{\eta}_{N}  
                 \; ,                     
\end{eqnarray} 
is realized in Eq. (\ref{DVN}) by the generalized, ${\cal O }_N$, Wigner
$D$ functions, where $g_{A,A-1}$ is the group element that corresponds to
the rotation (\ref{rotation2}), i.e. a rotation by an angle 
$\gamma=\arcsin(\sqrt{\frac{A-2}{2(A-1)}})$ in the 
$\vec{\eta}_{N},\vec{\eta}_{N-1}$ plane.

Since the two--body operator in Eq. (\ref{DVN}) depends only on the 
coordinate $ \vec{\xi}_{N}$ we should 
separate the hyperspherical functions related to this coordinate 
in the bra and the ket states. So, the last step in the derivation
consists of using the ohscfps introduced in 
Eq. (\ref{cfpON}). Then, the last term on the rhs of Eq.  (\ref{DVN}) is 
\begin{eqnarray} \label{lasteq}
&  &     < K_{N} L_{N} M_{N} \mbox {\boldmath $\lambda$}_{N} \beta^K_{N}   
    \tilde{\bf \Lambda}^{\prime \prime} _{N-1} 
    Y_{A-2}  \beta^{\lambda}_{A-2}) 
    |  O^R (\sqrt{2} \vec{\xi}_{N} ) 
    | K'_{N} L'_{N} M'_{N} \mbox {\boldmath $\lambda$}'_{N} 
    \beta^{K \, \prime}_{N} \tilde{\bf \Lambda}^{ \prime \prime}_{N-1} 
    Y_{A-2}  \beta^{\lambda\,\prime}_{A-2}) > =
\nonumber \\ & & 
   \sum_{K_{N-1} L_{N-1} \beta^K_{N-1}  \ell_{N}, \ell'_{N} } 
  \left[  \big( K_{N-1} L_{N-1} \mbox {\boldmath $\lambda$}_{N-1} 
  \beta^K_{N-1};\ell_{N} \big)  K_{N} L_{N} | \} 
  K_{N} L_{N} \mbox {\boldmath $\lambda$}_{N} \beta^K_{N} \right]
  \nonumber \\ & & \hskip 2truecm
\times \left[  \big( K_{N-1} L_{N-1} \mbox {\boldmath $\lambda$}_{N-1} 
  \beta^K_{N-1};\ell'_{N} \big)  K'_{N} L'_{N} | \} 
  K'_{N} L'_{N} \mbox {\boldmath $\lambda$}'_{N} \beta^{K \, \prime}_{N} 
  \right]
  \nonumber \\ & & \hskip 1truecm \times 
  < \big( K_{N-1} L_{N-1} \ldots; \ell_{N} \big) K_{N} L_{N} M_{N}  |
         O^R(\sqrt{2}\vec{\xi}_{N}) 
  | \big( K_{N-1} L_{N-1} \ldots  ; \ell'_{N} \big)
    K'_{N} L'_{N} M'_{N}  > \; \; .
\end{eqnarray}
This matrix element can be calculated for any given two--body operator. 
For scalar operator $O^R(r_{A,A-1})$, one obtains
\begin{eqnarray} \label{scalar2bme}
\lefteqn{ O^R_{K_{N} K'_{N}; K_{N-1} \ell_N}(\rho) \equiv }
    \nonumber \\ & & 
  < \big( K_{N-1} L_{N-1} \ldots; \ell_{N} \big) K_{N} L_{N} M_{N}  |
         O^R(\sqrt{2}{\xi}_{N}) 
  | \big( K_{N-1} L_{N-1} \ldots  ; \ell'_{N} \big)
    K'_{N} L'_{N} M'_{N}  > =
    \cr & & \hskip 1truecm
    \delta_{\ell_N, \ell'_N}
    {\cal N}_{N}( K_{N} ; \ell_{N} K_{N-1} )  
    {\cal N}_{N}( K'_{N} ; \ell_{N} K_{N-1} )  
  \; \int _0^{\frac{\pi}{2}} d \theta 
  \sin^{2 \ell_{N}+2}(\theta) \cos^{2 K_{N-1}+3N-4}(\theta)
\cr & & \hskip 2truecm \times
  P_{n}
   ^{(\ell_{N}+\frac{1}{2},K_{N-1}+\frac{3N-5}{2}) }(\cos{2\theta} )
  P_{n'}
   ^{(\ell_{N}+\frac{1}{2},K_{N-1}+\frac{3N-5}{2}) }(\cos{2\theta} ) 
   O^R(\sqrt{2} \rho \sin \theta)  \, ,
\end{eqnarray} 
with $n=\frac{K_N-\ell_N-K_{N-1}}{2}$,
$n'=\frac{K'_N-\ell_N-K_{N-1}}{2}$ and where
\begin{equation} \label{norN}
  \bbox {\cal N}_N(K_N ; \ell_N K_{N-1}) = \left[ \frac
  {(2K_N+3N-2)n!\Gamma(n+K_{N-1}+\ell_N+\frac{3N-2}{2})}
  {\Gamma(n+\ell_N+\frac{3}{2})
  \Gamma(n+K_{N-1}+\frac{3N-3}{2})} 
  \right]^{\frac{1}{2} } 
\end{equation}
is a normalization constant, and $P_n^{(a,b)}$ are the Jacobi polynomials.

The spin--isospin part of the two--body operator is calculated in a similar
way, first we use the last two generations of the spin--isospin cfps (
Eq. (\ref{stcfps})) to get the explicit dependence of the state 
$| S_A {S^z}_A T_A {T^z}_A \, \tilde{\Gamma}_{A} \tilde{Y}_{A-1}
        \, \alpha^{ST}_A >$
on the spin and the isospin of the last two--particles,
\begin{eqnarray} \label{stme1}
& &< S_A {S^z}_A T_A {T^z}_A \,{\Gamma}_{A} {Y}_{A-1} \, \alpha^{ST}_A | 
    O^{ST}(s_A,s_{A-1},t_A,t_{A-1})    
   | S'_A {S'^z}_A T'_A {T'^z}_A \,{\Gamma}'_{A} {Y}'_{A-1}\,\alpha'^{ST}_A >
   = \cr & & \hskip 1truecm
   \delta_{Y_{A-2}, Y'_{A-2}}
   \sum_{S_{A-1}T_{A-1}\alpha^{ST}_{A-1}}
   \sum_{S'_{A-1}T'_{A-1}\alpha'^{ST}_{A-1}}
   [ (S_{A-1}; s_A)S_A (T_{A-1} ; t_A)T_A \Gamma_{A-1}\alpha^{ST}_{A-1}
   |\} S_A T_A {\Gamma}_{A} \alpha^{ST}_A ]
   \cr & & \hskip 1truecm \times
   [(S'_{A-1}; s_A)S'_A (T'_{A-1} ; t_A)T'_A \Gamma'_{A-1}\alpha'^{ST}_{A-1}
   |\} S'_A T'_A {\Gamma}'_{A} \alpha'^{ST}_A ]
   \cr & & \hskip 1truecm \times
   \sum_{S_{A-2}T_{A-2}\alpha^{ST}_{A-2}}
   [ (S_{A-2}; s_A)S_{A-1} (T_{A-2} ; t_A)T_{A-1}\Gamma_{A-2}
   \alpha^{ST}_{A-2} |\} S_{A-1} T_{A-1} {\Gamma}_{A-1} \alpha^{ST}_{A-1} ]
   \cr & & \hskip 1truecm \times
   [ (S_{A-2}; s_A)S'_{A-1} (T_{A-2} ; t_A)T'_{A-1}\Gamma_{A-2}
   \alpha^{ST}_{A-2} |\} S'_{A-1} T'_{A-1} {\Gamma}'_{A-1} \alpha'^{ST}_{A-1} ]
   \\ & & \hskip 1truecm \times
   < S_A T_A S_{A-1} T_{A-1} S_{A-2} T_{A-2} \ldots |
   O^{ST}(s_A,s_{A-1},t_A,t_{A-1})  
   | S'_A T'_A S'_{A-1} T'_{A-1} S'_{A-2} T'_{A-2} \ldots > \nonumber
   \,.
\end{eqnarray}
Then the matrix element on the rhs of (\ref{stme1}) is calculated
using the $6j$ symbols.

Special attention should be payed to the matrix elements of the generalized
angular momentum operator $\hat{K}$ as it contains the hyperangular part
of the Laplace operator. Using the definition in Eq. (\ref{defME}) we
see that
\begin{equation}
  T_{K'\nu'p',K\nu p} =
        < K' \nu' | \chi_{p'}(z_{ij})^* \hat K \chi_p(z_{ij})
    | K \nu > \, .
\end{equation}
This matrix element can be easily calculated once we know the action
of $\hat K$ on $\chi_p(z_{ij})$. Since $\hat K$ is invariant under
kinematical rotations we can always write it using a set of Jacobi coordinates 
such that $\vec\xi_N=\frac{1}{\sqrt{2}}(\vec{ r}_i-\vec{ r}_j)$, which leads
to the following expression 
\begin{eqnarray}
& &< K' \nu' | \chi_{p'}(z_{ij})^* \hat K \chi_p(z_{ij}) | K \nu > = 
   \cr & & \hskip 2truecm
   < K' \nu' | \chi_{p'}(z_{ij})^* K(K+3N-2) \chi_p(z_{ij}) 
    - \chi_{p'}(z_{ij})^* (1-z^2_{ij})\frac{d^2 \chi_p(z_{ij})}{d z_{ij}^2}
   \cr & & \hskip 3truecm
    + \chi_{p'}(z_{ij})^* \{ z_{ij} + \frac{-2+(3N-2)z^2_{ij}}{z_{ij}} \} 
      \frac{d \chi_p(z_{ij})}{d z_{ij}}
   \cr & & \hskip 3truecm
    - 2 \chi_{p'}(z_{ij})^* 
     (1-z^2_{ij})\frac{d \chi_p(z_{ij})}{d z_{ij}} \frac{d}{d z_{ij}}
     | K \nu > \,.
\end{eqnarray}
This is just the matrix element of a sum of two body operators which can 
be calculated as explained above.

A bit more problematic are the three-- and four--body integrals 
$W^{[3]}_{K'\nu'\,p'\, K \nu\,p}$ and 
$W^{[4]}_{K'\nu'\,p' K\nu\,p}$. In order to calculate the three-- (four--)
body term we must first use three (four) generations of the ocfps in order
to express the matrix elements in terms of the orthogonal symmetry
adapted hyperspherical functions. Then we can use the proper rotations
and reduce the integrals to six--dimensional integrals that depend
only on the last two Jacobi coordinates. The explicit dependence of
the HH functions on the Jacobi coordinates is then revealed using the
last two generations of the ohscfps, Eq. (\ref{cfpON}). 
For $L=0$ states, these
integrals can further be reduced to three--dimensional integrations.


\section{DISCUSSION OF THE RESULTS}\label{results}

The present method becomes more and more complex as the number 
of fermions increases. Therefore in this work we consider 
only central $NN$ potentials. We present results
for 6, 8 and 12 nucleons interacting via the Volkov ~\cite{Volkov} (VV),
the Afnan--Tang ~\cite{Afnan} (S3), the modified S3 potential 
~\cite{ModAfnan} (MS3), the Brink--Boeker ~\cite{Brink} (B1) 
and the Malfliet--Tjon ~\cite{Malfliet} potentials (MT--I/III and MT--V).
The accuracy and convergence of the method have been investigated
for 6 and 8 nucleons.

The first step in our numerical study was to determine the
hyperradial working point, i.e. the parameters $a$ and $b$ of the Laguerre
polynomials in Eq. (\ref{Frho}), the number of radial grid points and 
their location.
After some preliminary tests we decided to use 15 hyperradial grid
points and 15 hyperradial functions  
with the parameters $b=0.25$ fm and $a=12$. The grid points
where chosen as the abscissas for a Gauss--Laguerre integration.

\begin{figure}[p]
\vspace{16.0truecm}
\includegraphics{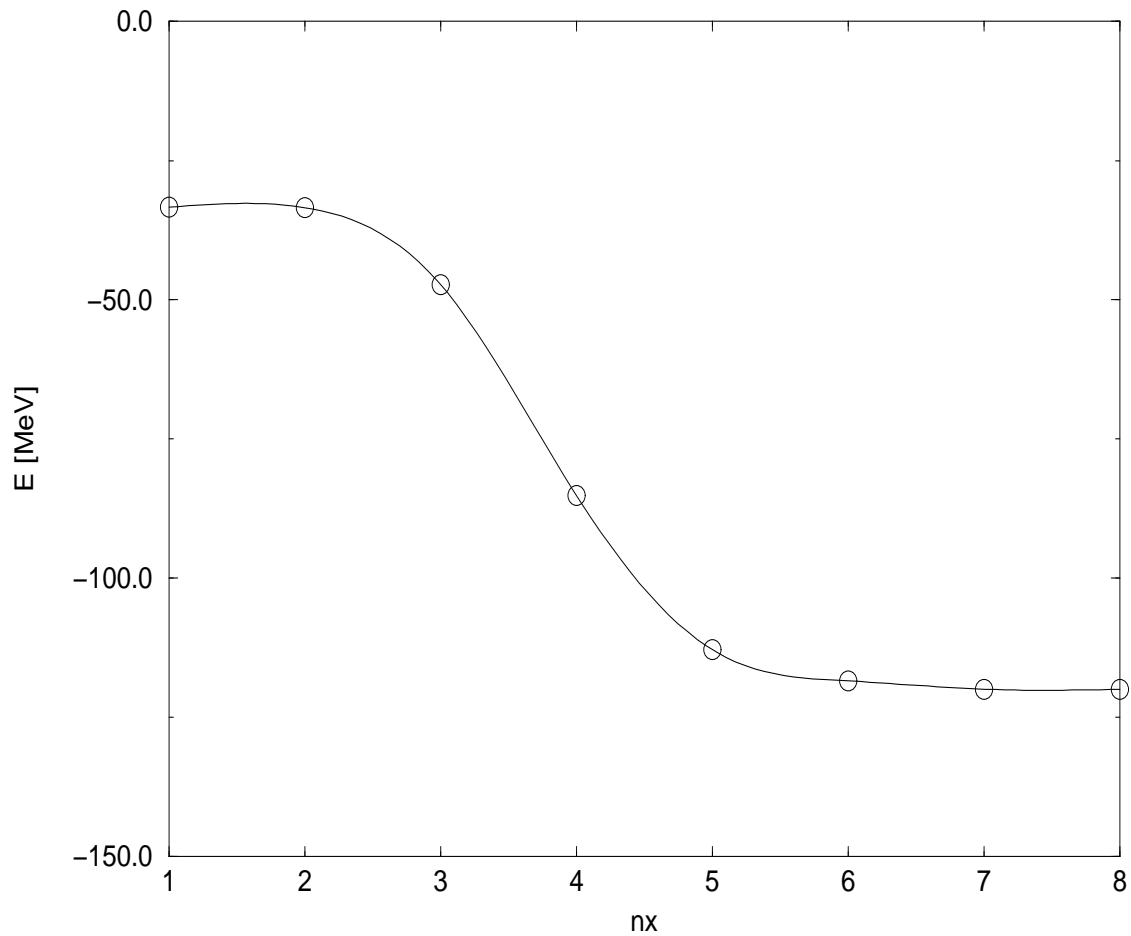}
\caption{Convergence of the binding energy with the number $n_x$ of polynomials
in the expansion of the correlation function.
Results are presented for $A=8$ and MTV potential.}
\label{fig:corr}
\end{figure}

The next step in setting our working point was to check the convergence
of the two--body correlation terms. As an example we studied the $^8$Be
system with the hard core MT--V potential, using a single
HH function, with $K=4$ and expanding the correlation function into 
Jacobi polynomials. 
As can be seen from Fig. \ref{fig:corr} the binding energy converges
very fast with increasing number $n_x$ of polynomials . 
In fact, the energy difference between 
the calculations with 7 and 8 correlations terms is about 0.02 MeV.
As a consequence we used 8 terms in the expansion of the correlation 
function.

The effect of introducing the hypercentral potential $V_{hc}(\rho)$ is
presented in Table \ref{tb:Vhc} for $^6$Li with the VV
 potential. As can be seen the hypercentral potential accelerates
the convergence of the HH expansion. With the lowest $K_{max}$ of 2
one already obtains a larger binding energy than with $K_{max}=6$
without hypercentral potential. Thus the introduction of $V_{hc}$
is certainly advantageous.

Our numerical results with the parameters described above are presented 
in Table \ref{tb:n6} for the $6$ nucleon system and in Table \ref{tb:n8}
for the $8$ nucleon system.

For $^6$Li the calculations include only one line
irreps of the kinematical group ${\cal O}_5$ and 
the irreps [42] and [33] of the permutation group. 
We compare our values with recent accurate variational
results  ~\cite{Varga95} available for some of these central potentials.
One can notice that with the VV potential one reaches convergence 
faster than with other potentials. The result for the binding energy 
starts oscillating around the asymptotic value. It also compares nicely
to the variational result of Ref. ~\cite{Varga95}. The other potentials 
show a tendency to convergence even if $K_{max}$ is not large enough to 
reach it. The MT potentials seem to lead to a more rapid convergence
than S3 and B1. The differences may be due
both to the fact that the convergent value has not yet been reached and
to the missing irreps of the permutation group.
The two line irreps of the orthogonal group are of little importance.
This has been checked for
the 8 particle case, where they give rather small contributions.

The calculations for $^8$Be 
include only the irrep (400) of the kinematical group ${\cal O}_7$  and 
irreps with at most 3 rows of the permutation group. Here the
comparison is made with the TICI results ~\cite{GMNBPW96}. In all cases our
results for the binding energy are somewhat larger. They
show characteristics similar to the six--body case. Again one sees
that the VV potential
result presents small oscillations around the convergent value and 
that the rather hard core MT potentials  seem to give
values closer to convergence than B1 or S3.
From the comparison between the TICI results with and without state dependent
correlations one can infer that for the MT potentials state independent 
correlations already give rather satisfying results, while
state dependent correlations are more effective for 
B1 and S3. Since our correlations are state independent one could expect such
a different convergence behavior as found in Table  \ref{tb:n8}

The results for $^{12}$C are presented in Table \ref{tb:n12}.
Here the calculations include a single HH state, $K_{max}=8$, 
with the kinematical group ${\cal O}_{11}$
irrep $(4,4,0)$ and the ${\cal S}_{12}$ irrep $[444]$.
The comparison is made with TICI ~\cite{GMNBPW96}, IDEA ~\cite{Brizzi96},
and the variational result of Ref. ~\cite{BBG87}.
Again from the comparison with the TICI results one can see that S3 and B1
results are farer from convergence than the MT values and that 
state dependent correlations play a similar role as discussed 
for the $^8$Be case. 

The IDEA results for 12 particles deviate somewhat from ours, although
they are all obtained using a single HH state. One possible explanation of 
the difference can be attributed to the choice of the HH state which
is not unique.
On the other hand we have obtained excellent agreement with the IDEA results
for 16 bosons.

\section{CONCLUSIONS}\label{concl}

In this work we have formulated a general method  
to calculate the wave functions of light systems up to considerably
large number of particle. This method combines the main ideas of
the HH expansion, the pair correlation ansatz and the Faddeev approach.
The actual application of it is made possible by the use of a
very efficient recursive algorithm to construct the antisymmetric A-particle
state containing hyperspherical harmonics. We have applied the method to
calculate the binding energies of 6, 8 and 12 nucleon systems with 
central local potentials.
The results we have obtained are very encouraging. For some potentials
(VV, 6 particles and MTV, 8 particles) we have reached the 
convergence region with $K_{max}=8$ which is the maximum value allowed 
by our present computer facilities (workstations).
The $^6$Li result for the VV potential is slightly higher than the SVM result.
Even if in other cases we  have not yet reached the convergence in the HH expansion our results are close to the TICI results. For the eight--body case
they are higher for all the potentials where results were
available for a comparison.

The method  presented here for the solution of the few--body Schr\"odinger
equation can be easily extended to solve the Schr\"odinger--like 
equation with a source, necessary for the application of the Lorentz integral
transform method. Work in this direction is in progress. 
\begin{table} 
\caption{\label{tb:Vhc} \em \\
         The effect of the hypercentral potential $V_{hc}$ on the 
         binding energy of six--nucleon system $^6$Li,
         $(L,S)J^{\pi}=(0,1)1^{+}$, interacting via the VV
         potential. }
\begin{center}
\begin{tabular}{lcc} 
  $K_{max}$ & \hspace{6mm} $V_{hc}=0$    \hspace{6mm}
	    & \hspace{6mm} 
         $V_{hc}=\int d \Omega \, H_{K_{min} \nu}^{\dagger} 
                 V_{1,2} H_{K_{min}\nu} $  \hspace{6mm} \\ \hline 
 2 & 64.18 & 66.10  \\
 4 & 64.81 & 66.53  \\
 6 & 65.47 & 66.63  \\
\end{tabular}
\end{center}
\end{table}

\begin{table} 
\caption{\label{tb:n6} \em \\
         Binding energies of six--nucleon system $^6$Li,
         $(L,S)J^{\pi}=(0,1)1^{+}$, interacting via various 
         NN potentials. $N_{HH}$ represents the number of hyperspherical
         harmonic states.}
\begin{center}
\begin{tabular}{lcccccc} 
  $K_{max}$ & $N_{HH}$ & \hspace{6mm} B1       \hspace{6mm}
		       & \hspace{6mm} MT--I/III \hspace{6mm}
		       & \hspace{6mm} MTV      \hspace{6mm}
		       & \hspace{6mm} S3       \hspace{6mm}
                       & \hspace{6mm} VV   \hspace{6mm} \\ \hline 
 2 &  1  & 30.99 & 30.15 & 62.45 & 62.76 & 66.10  \\
 4 &  4  & 37.82 & 34.67 & 63.27 & 64.45 & 66.53  \\
 6 & 12  & 39.11 & 35.43 & 64.10 & 66.49 & 66.63  \\
 8 & 31  & 39.61 & 35.91 & 64.55 & 67.18 & 66.57  \\
\hline
SVM \cite{Varga95}  &      
         &   -    &  -    & 66.30 & 70.65 & 66.25  \\ 
\end{tabular}
\end{center}
\end{table}

\begin{table}
\caption{\label{tb:n8} \em \\
         Binding energies of eight--nucleon system $^8$Be,
         $(L,S)J^{\pi}=(0,0)0^{+}$, interacting via various 
         NN potentials. Also given are results from Ref.
         \protect\cite{GMNBPW96} with state independent TICI$_{SI}$ 
         and state dependent TICI $_{SD}$ correlations. }
\begin{center}
\begin{tabular}{lcccccc} 
  $K_{max}$ & $N_{HH}$ & \hspace{6mm} B1       \hspace{6mm}
		       & \hspace{6mm} MT--I/III \hspace{6mm}
		       & \hspace{6mm} MTV      \hspace{6mm}
		       & \hspace{6mm} S3      \hspace{6mm}
                       & \hspace{6mm} VV   \hspace{6mm} \\ \hline 
 4 &  1  & 56.71 & 52.82 & 134.29  & 31.19 & 147.42  \\
 6 &  4  & 65.39 & 59.31 & 137.72  & 38.08 & 148.70  \\
 8 & 15  & 70.03 & 60.64 & 137.80  & 42.11 & 148.49  \\
\hline
TICI$_{SI}$  &      
         & 49.18 & 46.67 & 129.25  & 26.26 &         \\ 
\hline
TICI$_{SD}$  &      
         & 61.30 & 52.67 & 130.23  & 37.30 &         \\ 
\end{tabular}
\end{center}
\end{table}

\begin{table}
\caption{ \label{tb:n12} \em \\
         Binding energy of twelve nucleon system $^{12}$C,
         $(L,S)J^\pi=(0,0)0^+$, interacting via various 
         NN potentials. Also given are results from Refs.
         \protect\cite{GMNBPW96} (TICI),\protect\cite{Brizzi96} 
         (IDEA) and \protect\cite{BBG87} (VMC). 
         The TICI results as in Table \ref{tb:n8}. }

\begin{center}
\begin{tabular}{lccccc} 
                  & \hspace{6mm} B1       \hspace{6mm}
	          & \hspace{6mm} MT--I/III \hspace{6mm}
	          & \hspace{6mm} MTV      \hspace{6mm}
	          & \hspace{6mm} S3      \hspace{6mm}
                  & \hspace{6mm} VV   \hspace{6mm} \\ \hline 
This work               &  96.64 & 108.34 & 437.25 & 52.42 & 494.00  \\
TICI   & 103.93 & 109.04 & 429.44 & 62.99 &         \\
IDEA   &  80.1  &        &        & 44.4  &         \\
VMC    &  82.9 $\pm$ 0.2 &  &     &       &         
\end{tabular}
\end{center}
\end{table}

\subsection*{Acknowledgments}
The authors would like to thank V.D. Efros, V.B. Mandelzweig and R. Guardiola
for stimulating discussions.

\newpage


\end{document}